\documentclass[runningheads]{llncs}  
\usepackage{amsmath,amsfonts,amssymb}
\usepackage{graphicx}

\usepackage{subfigure}
\usepackage[subfigure]{tocloft}

\usepackage{amsmath,graphicx}
\usepackage{array}
\usepackage{lipsum}
\usepackage{xcolor}
\usepackage{float}
\usepackage{subfigure}
\usepackage{algorithm}
\usepackage{algorithmic}
\usepackage{lineno}
\linenumbers
\graphicspath{ {./images/} }
\usepackage{enumitem}
\setlist{nosep}

\usepackage{mwe} 

\title{De-speckling of Optical Coherence Tomography Images Using Anscombe Transform and a Noisier2noise Model}

\title{De-speckling of Optical Coherence Tomography Images Using Anscombe Transform and a Noisier2noise Model}
\titlerunning{De-speckling and Using Anscombe Transform and a Noise2noise Model}

\author{Arka Saha\inst{1} \and
Sourya Sengupta\inst{2}} 

\institute{Dept. of Electrical Engineering, Jadavpur University, India \and
Dept. of Electrical and Computer Engineering, University of Illinois, Urbana-Champaign, Illinois, USA \\}

\cftpagenumbersoff{figure}
\cftpagenumbersoff{table} 
\begin{document} 
\nolinenumbers
\maketitle
\begin{abstract}
Optical Coherence Tomography (OCT) image denoising is a fundamental problem as OCT images suffer from multiplicative speckle noise, resulting in poor visibility of retinal layers. The traditional denoising methods consider specific statistical properties of the noise, which are not always known. Furthermore, recent deep learning-based denoising methods require paired noisy and clean images, which are often difficult to obtain, especially medical images. Noise2Noise family architectures are generally proposed to overcome this issue by learning without noisy-clean image pairs. However, for that, multiple noisy observations from a single image are typically needed. Also, sometimes the experiments are demonstrated by simulating noises on clean synthetic images, which is not a realistic scenario. This work shows how a single real-world noisy observation of each image can be used to train a denoising network. Along with a theoretical understanding, our algorithm is experimentally validated using a publicly available OCT image dataset. Our approach incorporates Anscombe transform to convert the multiplicative noise model to additive Gaussian noise to make it suitable for OCT images. The quantitative results show that this method can outperform several other methods where a single noisy observation of an image is needed for denoising. The code and implementation of this paper will be available publicly upon acceptance of this paper.

\end{abstract}


\section{Introduction}
Optical Coherence Tomography (OCT) imaging technique is a widely popular imaging modality in ophthalmic diagnosis due to its non-invasive nature \cite{agarwal2015essentials}. OCT images can provide cross-sectional micro-structure views of biological tissues. OCT imaging is based on the interferometric technique, initially invented by Albert A. Michelson. OCT is used to study different retinal layers to diagnose important retinal diseases, such as glaucoma, Age-related macular degeneration (AMD), diabetic retinopathy (DR), etc. A more detailed review of information and applications about OCT can be found at \cite{huang1991optical}
Laser speckle noise is always present in optical coherence tomography (OCT) images. The speckle occurs due to the coherent nature of the light sources \cite{sengupta2020deep}. Due to speckle, the image quality, contrast, and signal-to-noise ratio get poor. Moreover, having clear visibility of retinal layers is essential for different diagnosis, the speckle results in the reduction of the visibility of the retinal layers and the tissues. Hence, it is important to denoise the OCT images. 

\section{Related Work}
In the existing literature, different denoising methods like adaptive median filtering, linear least square estimation , wavelet-based methods , 3D filtering (BM3D), dictionary learning (k-SVD) have been proposed to denoise the OCT images.  \cite{sengupta2020deep}. But, these methods hold specific assumptions about the noise's statistical properties, and hence, the application of these techniques is limited. Recently, deep learning methods have become popular for solving different denoising problems. These methods use noisy-clean image pairs for training a supervised network. Though these models have performed better than traditional methods, given that the dataset is large enough, they need paired noisy and clean images to train the model. This is often rare to find in real-life scenarios. To overcome this issue, recently, a family of methods, like Noise2Noise \cite{lehtinen2018noise2noise}, Noisier2noise \cite{moran2020noisier2noise} have been proposed. The authors have shown that noisy input-target pairs can also be used to perform denoising. In a recent work, Guillame et al. \cite{gisbert2020self} used Noise2Noise family network for image registration and denoising of OCT data. Bin et al. \cite{qiu2021comparative} compared performances of different CNN architectures in Noise2Noise methods to denoise OCT images. However, these above-mentioned methods used multiple noisy observations of an image. This approach is time-consuming and unrealistic to acquire many noisy images from a particular patient. Huang et al. \cite{huang2021real} required multiple noisy observations for a single image and also assumed an additive noise model, whereas OCT suffers from multiplicative speckle noise. 
\begin{figure*}[t]
    \centering
    \subfigure[]{\includegraphics[width=0.24\textwidth]{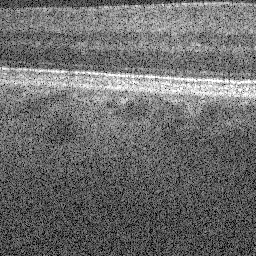}}
    \subfigure[]{\includegraphics[width=0.24\textwidth]{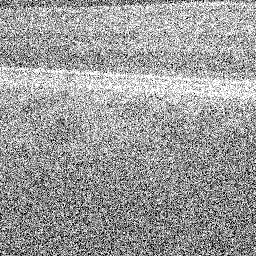}} 
    \subfigure[]{\includegraphics[width=0.24\textwidth]{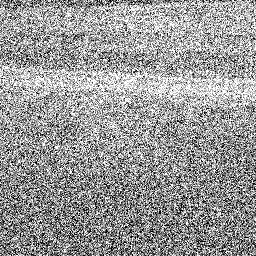}} 
    \subfigure[]{\includegraphics[width=0.24\textwidth]{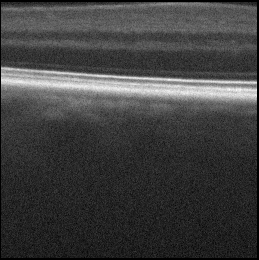}}
    \caption{Fig: From left to right: (a) Original noisy (b) Noisier (c) Noisier+, (d) Clean image
 }
    \label{fig:foobar}
\end{figure*}
In order to circumvent these, availability of a single noisy observation for an image is considered in this method.\\
Specifically, the significant contributions of our work are \\ 1. Unlike previously published Noise2Noise family methods on OCT image denoising, this approach does not need more than one real-noisy observation of a particular image. All the simulations and training were done upon one single noisy observation per image. \\
2. Our approach theoretically establishes the desired outcome and experimentally validates that.

\section{Methods}
We are only considering that a single noisy observation $X_{noisy}$ of a particular image $X$ is available. Now to train the model, two noisier versions are generated from that single image, namely $Z$ and $Y$ such that: 
$Z = X_{noisy} + M_{1} + M_{2}$ and $Y = X_{noisy} + M_{1}$
Throughout the rest of the paper, $X_{noisy}$, $Y$, and $Z$ are termed as real-noisy, noisier, noisier+ respectively. Here $M_{1}$ and $M_{2}$ are two zero-mean additive Gaussian noise, and it can be safely assumed that the mean and variance of the original noise level in the noisy observation is much smaller than the signal intensity. Hence it can be written that: $E[n_{0}] << E[X]$. Following these notations and assumptions:
\begin{equation}
\begin{split}
    E[Y|Z] = E[X_{noisy} + M_{1} | Z]  \\
            = E[X_{noisy}|Z] + E[M1|Z] \\ 
            = E[X_{noisy}|Z] \\
            \approx E[X|Z]
\end{split}
\end{equation}

{Hence by estimating the less noisy counterpart of the noisier image, the estimate of the clean image can be obtained. It is to be noted that there can be many possible clean images corresponding to a noisy image, and this method will only generate a mean estimate of the possible solutions. A similar level of uncertainty holds even with a fully supervised method with noisy-clean paired images.}\\
The process mentioned in the algorithm assumes the noise model as additive, whereas in the case of OCT images, the noise model is generally modeled as multiplicative Generalized Gamma distribution \cite{kirillin2014speckle}.
\begin{algorithm}[H]
\begin{algorithmic} 
\STATE \textbf{Input:} input real-noisy image patches ($X_{noisy}$)
\STATE \textbf{Step 1:} Anscombe transformation of real-noisy image patches ($X_{ans}$)
\STATE \textbf{Step 2:} Noisier version ($Y$) of the ($X_{ans}$). 
\STATE \textbf{Step 3:} Noisier version ($Z$- termed as noisier+) of the ($X_{ans}$).
\STATE \textbf{Step 4:} Initialize the UNet image-to-image translation network.
\STATE \textbf{Step 5:} Fix training parameters, such as: learning rate, optimizer, batch size.
\WHILE{stopping criteria is not met}
\STATE Train the network to learn a mapping from noisier+ ($Z$) image to noisier image ($Y$).
\ENDWHILE
\end{algorithmic}
\end{algorithm}
To make it suitable for the proposed approach, the Anscombe transform \cite{makitalo2010optimal} was used to convert multiplicative noise to additive noise. 
This operation approximates the multiplicative noise of the OCT image to an additive Gaussian noise with unit variance. The network was trained using Anscombe transformed versions of noisier+ and noisier image patches. The quality of the inverse Anscombe output of the final denoised image was measured with respect to the original ground truth images during inference. 
Experimentally, we found that during inference, passing the denoised output image multiple times through the network enhanced the denoising effect; Hence the output image of the trained network was again fed to the trained network and the process was repeated multiple times. The final metrics were calculated after passing the denoised output image 2 times through the network. There was no significant improvement in quality after 2 times. The steps of the proposed algorithm are described in Algorithm 1.
\section{Numerical Studies}
\subsection{Dataset}
The DUKE SD-OCT image dataset \cite{fang2013fast}. was used in this study. The dataset has 28 images with a resolution of 450X900. A Biopitgen SD-OCT imaging system was used to capture the images. Several noisy B-scan images, obtained at the same position, were registered and averaged to get a corresponding clean image. The noisier version of the image was simulated by adding additive Gaussian noise to the Anscombe transformed original noisy observation once, the noisier+ version of the image was simulated by adding additive Gaussian noise to the simulated noisier observation. {The standard deviation of the Gaussian noise used for all purposes was kept at 50.}
\subsection{Training Details} 
The dataset comprised 1700 OCT image patches, which were split into 1500 training patches, 100 validation patches, and 100 testing patches.
A U-Net \cite{ref:unet} style network was used as the backbone of the framework to train the model. Adam optimizer with a learning rate of 2e-5 and beta values of 0.9 and 0.99 was used for training. The model was trained with a batch size of 8 patches. The validation loss was monitored to fix the stopping rule during training, and the training was stopped when there was no decrease in validation loss for consecutive 4 epochs. Titan X Nvidia graphics processing unit (GPU) was used for all computations and experiments, codes were written on PyTorch framework.
\subsection{Results:}
\begin{figure*}[t]
    \centering
    \subfigure[Real Noisy Image]{\includegraphics[width=0.24\textwidth]{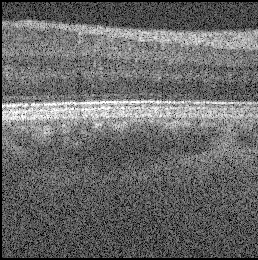}}
    \subfigure[Clean Counterpart]{\includegraphics[width=0.24\textwidth]{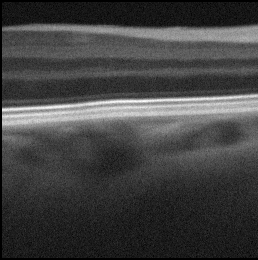}} 
    \subfigure[BM3D]{\includegraphics[width=0.24\textwidth]{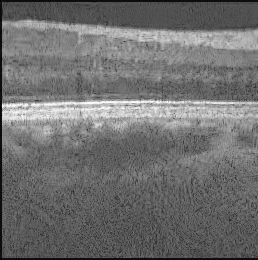}} 
    \subfigure[NLM]{\includegraphics[width=0.24\textwidth]{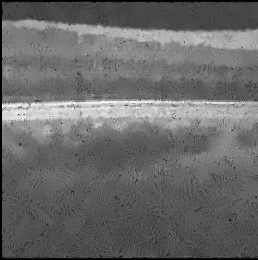}}
    \subfigure[TV]{\includegraphics[width=0.24\textwidth]{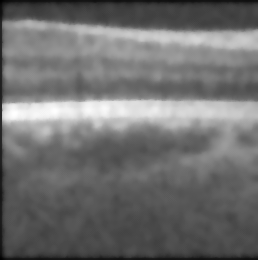}}
    \subfigure[Wavelet]{\includegraphics[width=0.24\textwidth]{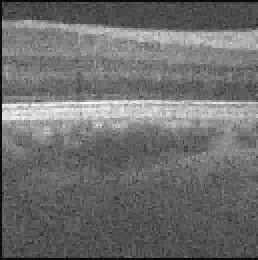}}
    \subfigure[  Ours]{\includegraphics[width=0.24\textwidth]{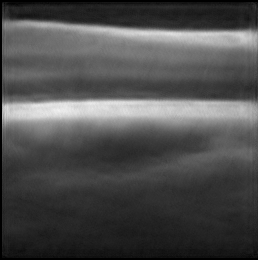}}
    \subfigure[  Supervised]{\includegraphics[width=0.24\textwidth]{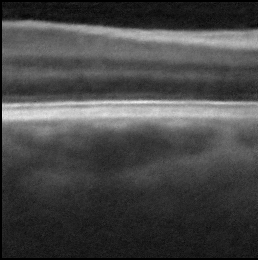}}
    \caption{Fig: (a) Noisy, (b) Ground truth, results of (c) BM3D (d) NLM (e) TV (f) Wavelet (g) Proposed method (h) Output of model trained via supervised learning. It can be seen that the proposed method has significantly lesser artifacts and reduced noises than the other comparable methods.
 }
    \label{fig:foobar}
\end{figure*}

\begin{table*}[t]
  \centering
  \begin{tabular}{|c|c|c|c|c|c|c|}
    \hline
    & TV \cite{chambolle2004algorithm} & Wavelet \cite{chang2000adaptive} & NLM \cite{buades2005non} & BM3D \cite{dabov2007image} &   Ours & Supervised\\
    \hline
  PSNR & 16.17 ± 1.61 & 16.03 ± 1.79 & 15.93 ± 1.76& 17.92 ± 1.81 & 21.86 ± 1.56 & 22.34 ± 1.65\\ 
  \hline
  SSIM & 0.76 ± 0.04 & 0.75 ± 0.07 & 0.75 ± 0.07 & 0.77 ± 0.09 & 0.87 ± 0.03 & 0.89 ± 0.08\\ 
  \hline

  \hline
  \end{tabular}
  \caption{Experiment results on different algorithms (mean ± standard deviation)}
  \label{tab:1}
\end{table*}
As this method does not require any clean counterpart of the real-noisy image, it is more suitable to compare with the denoising algorithms, which also do not require noisy-clean pair and can only operate on a single real-noisy observation of an image. Hence, 4 such widely-used denoising algorithms, namely: BM3D \cite{dabov2007image}, non-local means (NLM)\cite{buades2005non} , wavelet-based denoising \cite{chang2000adaptive}, and total-variation (TV) based denoising \cite{chambolle2004algorithm}, were compared with our model's output.  Table 1 contains the quantitative results in terms of mean and standard deviation of all different methods in tabulated form. The proposed method outperformed other existing approaches in terms of widely used image quality metrics PSNR and SSIM \cite{wang2004image}. For comparison and as an upper limit baseline, the results corresponding to noisy-clean paired image supervised training are also shown in the table. In Fig. 2 results of the methods are shown for visual comparison between the proposed method and other methods. Fig. 2 (a) is the original noisy image, (b) is the ground truth, (c-f) and (h) are the outputs given by traditional algorithms BM3D, NLM, TV-based denoising,and wavelet-denoising, paired training respectively. It can be observed that these methods may result in blurriness, low-level artifacts, and residual noises compared to the output of our proposed method, which is significantly cleaner in comparison. {The comparative baseline results with noisy-clean paired images are also shown in the same figure. Some more example results are shown in the attached supplementary file.}

\section{Conclusion}
In this work, we showed how a denoised image estimate could be obtained by just having a single real-life noisy observation of an image. Specifically, inspired by Noisier2Noise family methods of image denoising, we theoretically established a framework and provided experimental results with OCT images. The Anscombe transformation was used to redistribute the multiplicative Poisson noise as additive Gaussian noise. The experimental quantitative results signify that the framework surpassed other traditional denoising methods, which also need a single noisy image to estimate the denoised counterpart. {In the future work, the algorithm will be validated using other different datasets that involve multiplicative noise}
\section{Conflict of Interest}
The authors declare no conflict of interest.
\section{Acknowledgments}
This research does not involve any specific grant from any funding agency.

\bibliography{report}   
\bibliographystyle{spiejour} 

\end{document}